\documentclass[fleqn,12pt,twoside]{article}
\usepackage{espcrc1}

\usepackage{graphicx}

\newcommand{\AmS}{{\protect\the\textfont2
  A\kern-.1667em\lower.5ex\hbox{M}\kern-.125emS}}

\title{Carbon-Enhanced Metal-Poor Stars in the Early Galaxy}

\author{B. Marsteller, T.C. Beers, \address{Department of Physics \& Astronomy and JINA: Joint Institute for
        Nuclear Astrophysics, Michigan State University, USA} 
	S. Rossi, \address{Departamento de Astronomia, IAG, Universidade de S\~{a}o Paulo, SP, Brazil}
        N. Christlieb, \address{Hamburger Sternwarte, Universit\"at Hamburg, Germany} 
	M. Bessell, \address{Research School of Astronomy \& Astrophysics, Australian National University, Australia}
        and
	J. Rhee, \address{Center for Space Astrophysics, Yonsei Univ., Korea, and
		Space Astrophysics Laboratory, Caltech, USA}}
       
\begin{document}


\maketitle

\begin{abstract}

Very metal-deficient stars that exhibit enhancements of their carbon abundances
are of crucial importance for understanding a number of issues -- the nature of
stellar evolution among the first generations of stars, the shape of the Initial
Mass Function, and the relationship between carbon enhancement and
neutron-capture processes, in particular the astrophysical s-process. One recent
discovery from objective-prism surveys dedicated to the discovery of
metal-deficient stars is that the frequency of Carbon-Enhanced Metal-Poor (CEMP)
stars increases with declining metallicity, reaching roughly 25\% for [Fe/H] $<
-2.5$. 

In order to explore this phenomenon in greater detail we have obtained
medium-resolution (2 \AA) spectroscopy for about 350 of the 413 objects in the
Christlieb et al. catalog of carbon-rich stars, selected from the
Hamburg/ESO objective prism survey on the basis of their carbon-enhancement,
rather than metal deficiency. Based on these spectra, and near-IR $JHK$
photometry from the 2MASS Point Source Catalog, we obtain estimates of [Fe/H]
and [C/Fe] for most of the stars in this sample, along with reasonably accurate
determinations of their radial velocities. Of particular importance, we find that
the upper envelope of carbon enhancement observed for these stars is nearly
constant, at [C/H] $\sim -1.0$, over the metallicity range $-4.0 <$ [Fe/H] $<
-2.0$; this same level of [C/H] applies to the most iron-deficent star yet
discovered, HE 0107-5240, at [Fe/H] = -5.3.

\end{abstract}

\section{Introduction}

In recent studies of the most metal-poor stars known, an interesting trend has
been discovered regarding the abundance of carbon. Several authors have noted
that roughly 25\% of the most metal-poor stars studied exhibit strong enhancements
of carbon, as reflected in the unusual strength of their CH G-band features.
These Carbon-Enhanced Metal-Poor (CEMP) stars occur with such a large frequency
that one is forced to consider the implications on the nature of possible
production mechanisms of carbon in the early universe.

Over the metallicity range $-4.0 < $ [Fe/H] $ < -2.0$, there exists an upper
limit to the level of carbon enhancement amongst CEMP stars at [C/H] $\sim -1.0$
\cite{Ros99}. This immediately suggests that at some early time in
the universe a significant amount of carbon was produced, in all likelihood by
one of the following sources: (1) a {\it primordial} mechanism from massive
stellar progenitors, (2) {\it intrinsic} internal production by low-mass stars
of extremely low [Fe/H], or (3) {\it extrinsic} production of carbon by stars of
intermediate mass, which can be prodigious manufacturers of carbon during their
AGB stages, followed by mass transfer to a surviving lower-mass companion.
Indeed, it remains possible that all three sources might play a role.

The first alternative, in which the observed levels of carbon in at least some
of the CEMP stars is primordial, or close to primordial, and was produced in the
first generations of (presumably quite massive) stars, receives some support
from models of element production in zero metal abundance stars in the
mass range 18 -- 30 $M_{\odot}$ \cite{WW95}. In this scenario a large amount of carbon was
produced early on, and then not as prodigiously after that.

The second possibility is that, early in the universe, when there were few heavy
elements present, unusually effective mixing episodes triggered at the time of 
helium core flash dredges up internally produced carbon and deposits it on
the surfaces of low-mass stars \cite{Fuj00}.  

The final alternative is that intermediate-mass stars (e.g., $2 \le M_{\odot} \le
8$) in binary pairs with lower-mass companions, quickly evolved, producing large
amounts of carbon during their AGB evolution. Then, a significant amount of this
carbon-rich material was transferred to the long-lived companion, via roche-lobe
overflow or a wind. The lower-mass companion is presently observed to be
carbon-rich, while the higher-mass carbon-producing star is now a faint white
dwarf (see, e.g., \cite{Luc04}). In this scenario, one would expect
to be able to detect the presence of the binary system, either directly through
high-angular-resolution observations, or indirectly by detecting the tell-tale
wobble in the orbit of the visible companion.

\section{Further Investigations}

In order to investigate these different scenarios, new medium-resolution
spectroscopic data has been gathered, with a variety of 2m-4m class telescopes,
for over 350 stars in the large sample of carbon-rich stars reported by
Christlieb et al. \cite{Chr01}. It should be kept in mind that this sample of stars 
was selected based, not on metallicity, but rather, on the apparent level of
carbon enhancement revealed in low-resolution objective-prism specta. As a
result, only those metal-poor stars which exhibited enhancement of carbon were
selected, resulting in a determination of the upper limit of carbon enhancement,
without restricting the metallicity range. 

These data allow for the determination of reasonably accurate estimates of
carbon abundances (and [Fe/H]) for the Christlieb et al. carbon-rich stars,
using techniques described by Rossi et al. \cite{Ros04}. The calibration
of Rossi et al., based on the KP and GP indices described by Beers et al.
\cite{Bee99}, in combination with de-reddened $J-K$ colors obtained from the 2MASS
survey \cite{Cu03}, achieves accuracies of $\sigma = 0.25$ dex for [Fe/H] and $\sigma
= 0.30$ dex for [C/Fe], respectively.

Figure 1 shows the derived abundances that have been obtained to date for a
subset of the Christlieb et al. carbon-rich stars. Roughly 50\% of the
carbon-enhanced candidates proved to be very metal-deficient stars with [Fe/H]
$\le -2.0$, in other words, they are CEMP stars. The remainder appear to be
stars of higher metallicity, up to and including solar. These latter stars may
be examples of intermediate-mass intrinsic AGB stars, whose presence in the halo
of the Galaxy might well be accounted for by stripping from dwarf galaxies such
as Sagittarius \cite{Ib01}.

\begin{figure}[htb]
\begin{center}
\includegraphics[width=10cm, height=6cm]{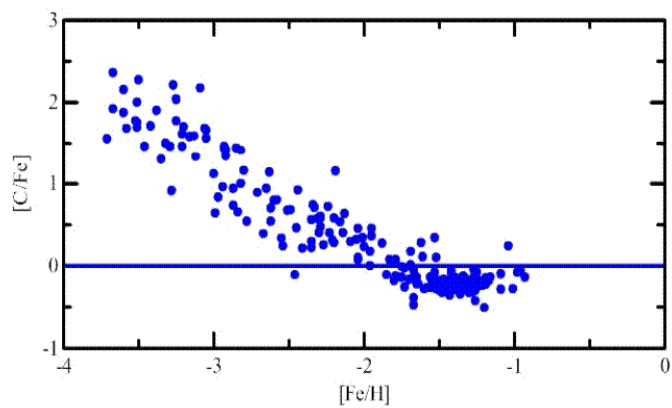}
\caption{Note the strong trend of increasing [C/Fe] with declining [Fe/H].
These stars were selected as carbon-enhanced, independent of metallicity. The
upper envelope of abundances correspond to [C/H] = $-1.0$.  Note that
abundances have not yet been determined for stars more metal-rich that [Fe/H] =
$-1.0$.}

\end{center}

\end{figure}

In addition to elemental abundances, radial velocities have been obtained for
most of these candidates. Around half of these stars have velocites
consistent with membership of the general halo population. The remainders are a
mix of high- and low-velocity stars, some of which are likely members of the
metal-weak thick disk \cite{Bee02}, while others are possibly associated with
the Sagittarius stream. 

High-resolution spectroscopy of the most interesting stars from this sample are
presently being obtained with a variety of 4m-8m telescopes, and will be
reported on in due course.

\end{document}